\documentclass[apj]{emulateapj}

\newcommand{\be}{\begin{eqnarray}}
\newcommand{\ee}{\end{eqnarray}}
\newcommand{\beq}{\begin{equation}}
\newcommand{\eeq}{\end{equation}}
\def\simless{\mathbin{\lower 3pt\hbox
      {$\rlap{\raise 5pt\hbox{$\char'074$}}\mathchar"7218$}}}
\def\simgreat{\mathbin{\lower 3pt\hbox
      {$\rlap{\raise 5pt\hbox{$\char'076$}}\mathchar"7218$}}} 

\def\Tdeep{T_{\rm deep}}

\begin{document}

\title{Thermal Structure and Radius Evolution of Irradiated Gas Giant
  Planets} 

\author{Phil Arras and Lars Bildsten}
\affil{ Kavli Institute for Theoretical Physics \\
Kohn Hall, University of California,
\\ Santa Barbara, CA 93106; arras@kitp.ucsb.edu, bildsten@kitp.ucsb.edu}

\begin{abstract}

We consider the thermal structure and radii of strongly irradiated gas
giant planets over a range in mass and irradiating flux. The cooling
rate of the planet is sensitive to the surface boundary condition,
which depends on the detailed manner in which starlight is absorbed
and energy redistributed by fluid motion. We parametrize these effects
by imposing an isothermal boundary condition $T \equiv T_{\rm deep}$
below the photosphere, and then constrain $T_{\rm deep}$ from the
observed masses and radii. We compute the  dependence of luminosity
and core temperature on mass, $T_{\rm deep}$ and core entropy, finding
that simple scalings apply over most of the relevant parameter space.
These scalings yield analytic cooling models which exhibit power-law
behavior in the observable age range $0.1-10\ {\rm Gyr}$, and are
confirmed by time-dependent cooling calculations. We compare our model
to the radii of observed transiting planets, and derive constraints on
$T_{\rm deep}$. Only HD 209458 has a sufficiently accurate radius
measurement that $T_{\rm deep}$ is tightly constrained; the lower error bar
on the radii for other planets is consistent with no irradiation. More 
accurate radius and age measurements will allow for a determination of
the correlation of $T_{\rm deep}$ with the equilibrium temperature,
informing us about both the greenhouse effect and day-night asymmetries.

\end{abstract}

\keywords{planetary systems--planets and satellites:general}


\section{Introduction}

\begin{deluxetable*}{lccccccr}
\tabletypesize{\small}
\tablewidth{0pt}
\tablecaption{Transiting Extrasolar Planets \label{tab1} }
\tablehead{
\colhead{object} & 
\colhead{$a$(au)} & 
\colhead{$M_{\rm p}$($M_{\rm J}$)} & 
\colhead{$R_{\rm p}$($R_{\rm J}$)} &
\colhead{ $T_{\rm eq}[K]$ \tablenotemark{a} } & 
\colhead{ $T_{\rm deep}[K]$\tablenotemark{b}} & 
\colhead{Age (Gyr)} & 
\colhead{Reference}  }
\startdata
  OGLE-TR-132  & 0.031 & $1.19 \pm 0.13$ & $1.13 \pm 0.08$ &
  2100& $\leq 2200$   & 0--1.4  & 1 \\
  OGLE-TR-56   & 0.023 & $1.24 \pm 0.13$ & $1.25 \pm 0.08$  & 2100
  & $ 1000-3100$ & $3 \pm 1$ & 2,3,12 \\
  HD 209458    & 0.046  & $0.69 \pm 0.05$  & $\rm
  1.31^{+0.05}_{-0.05}$ & 1500 & 2200-2800 & 4--7  & 4,5\\
  OGLE-TR-10   & 0.042  & $0.63 \pm 0.14$  & $1.14 \pm 0.09$   &
  1500 & $\leq 2600$  &  -- & 6,12  \\
  OGLE-TR-113 & 0.023 & $1.35 \pm 0.22$ & $\rm
  1.08^{+0.07}_{-0.05}$ & 1300 & $\leq 2100$ & -- &  7 \\
  TrES-1      & 0.039 & $0.73 \pm 0.04$ & $\rm
  1.08^{+0.05}_{-0.05}$    & 1200 & $\leq 1000$  & $2.5 \pm 1.5$ & 5,8 \\
  OGLE-TR-111  & 0.047 & $0.52 \pm 0.13$ & $\rm
  0.97 \pm 0.06$    &1000 & $\leq 1200$ & -- & 9,12 \\
  HD 149026 \tablenotemark{c}    & 0.042 & $0.36\pm 0.04$ & $0.725 \pm
  0.05$ & 1700 
  &  &  $2.0 \pm 0.8$  & 10 \\
  HD 189733 &  0.031 & $1.15\pm0.04$ & $1.26\pm 0.03$ & 1200 & $\leq 3200$ &
  &  11 \\ 
\enddata 
\tablenotetext{a}{Here $T_{\rm eq} \equiv T_{\rm
    *}(R_*/2a)^{1/2}$. See the discussion following eq.\
  (\ref{eq:Teq}).}
\tablenotetext{b}{Allowed range of $T_{\rm deep}$ given range of mass,
radius and age. If no age range given in the literature, we
(arbitrarily) give the maximum value of $T_{\rm deep}$ for an age less
than 10Gyr. However, given an accurate age range, the figures in
\S \ \ref{sec:applications} can be used to obtain stronger
constraints than given here.}
\tablenotetext{c}{HD 149026's small radius clearly indicates a large
  core size or heavy element abundance. The present paper does not
  include heavy element cores, so we do not discuss HD 149026 further.}
\tablerefs{(1) \cite{2004A&A...424L..31M}, (2)
  \cite{2004ApJ...609.1071T}, (3) \cite{2003ApJ...596.1327S},
  (4) \cite{2002ApJ...569..451C}, (5) \cite{2005ApJ...621.1072L},
  (6) \cite{2005ApJ...624..372K}, (7) \cite{2004A&A...421L..13B},
  (8) \cite{2004ApJ...616L.167S}, (9) \cite{2004A&A...426L..15P},
  (10) \cite{2005ApJ...633..465S},(11)\cite{2005astro.ph.10119B},
  (12) \cite{2006astro.ph..1024S} }
\end{deluxetable*}

Following the discovery of the planet orbiting 51 Peg
\citep{1995Natur.378..355M,1995AAS...187.7004M},  more than 160 planets
have been found around nearby stars
using precision Doppler spectroscopy.
\footnote{For up to date catalogs, see http://exoplanets.org/ and
http://obswww.unige.ch/~udry/planet/planet.html.} Theories of planet
formation now have the demanding task of explaining the existence of
gas giants with semi-major 
axes one hundred times smaller than Jupiter, others with order unity
orbital eccentricities, a detailed spectrum of (minimum) planet
masses, and metallicity correlations with the parent star.

The discovery of transiting planets in the last five
years (see Table \ref{tab1}) challenges not only theories for the
origin of short-period gas giants, but also their structure and
thermal evolution, spectrum, and interior fluid dynamics. Measurements
of planetary mass, radius, and (stellar) age test cooling models
which predict radius as a function of mass and age. The atmospheres of
two planets have been directly observed. For HD 209458b, absorption
lines (due to stellar photons  passing through planet's atmosphere)
have been found \citep{2002ApJ...568..377C, 2003Natur.422..143V,
2004ApJ...604L..69V}, and the first detections of photons
emitted by planets outside our solar system have been made for the
thermal emission from HD 209458b \citep{2005Natur.434..740D} and
TrES-1 \citep{2005ApJ...626..523C}. These observations directly
constrain the atmospheric structure, temperature profile and chemical
composition near the photosphere.

Evolution of the short orbital period transiting exoplanets is
significantly different than for Jupiter and Saturn due to proximity
of the parent star \citep{1996ApJ...459L..35G}. Irradiation increases
the photospheric temperature by  nearly an order of magnitude relative to
an isolated planet. Irradiation also decreases the cooling rate, and
hence the rate of shrinkage, by  altering the surface boundary
condition \citep{2000ApJ...534L..97B}. This is immediately apparent in
Table \ref{tab1} as many transiting extra-solar giant planets (EGP's)
have radii significantly larger than Jupiter. As short period planets
are expected to be tidally synchronized
\citep{1996ApJ...459L..35G,1997ApJ...481..926M}, 
the strong day-night 
temperature contrast will drive winds to transport heat from the day
to the night side \citep{2002A&A...385..166S}. Hence the atmospheric
temperature profile depends on a combination of detailed
radiative transfer calculations for absorption of starlight, and
hydrodynamics to model day-night winds and dissipation of wind kinetic
energy. Lastly, tides raised on the planet by the parent star may
significantly affect its thermal evolution
\citep{2001ApJ...548..466B,2002A&A...385..166S}. The  free energy
available by synchronizing the  planet's spin or circularizing the
orbit are comparable or larger than the thermal energy. Hence {\it if}
the heat can be deposited sufficiently deep in the planet in less than
a cooling timescale, the cooling can be slowed, or even reversed. 
However, it is uncertain if tides can deposit heat deep in
the planet \citep{1997ApJ...484..866L, 2004ApJ...610..477O,
2004astro.ph..7628W} .
  
Evolutionary models show that the cooling rate is quite sensitive to
the uncertain surface boundary condition
\citep{2002A&A...385..156G}. This boundary condition has been
implemented using various approximations. Full radiative transfer
calculations \citep{2001ApJ...556..885B, 2003ApJ...594.1011H} of {\it
static} atmospheres include the stellar irradiation self-consistently,
and determine the temperature structure for a given cooling flux from
the deep interior. These calculations compute (rather than assume) the
albedo, and determine the temperature rise due to absorption of
starlight (the greenhouse effect). Such detailed radiative transfer
solutions have been incorporated as boundary conditions for some
evolutionary calculations \citep{2003A&A...402..701B,
2003ApJ...594..545B}. However, as day-night and equator-pole winds are
not included, assumptions must be made about how the stellar flux is
deposited over the surface of the planet (only day-side versus evenly
over the entire surface, etc.) which directly affect the
temperature profile. Other evolutionary calculations
(e.g. Bodenheimer et al. 2003) solve the radiation diffusion
equation and set the temperature at (infrared) optical depth $2/3$ to
be the equilibrium temperature, ignoring additional temperature
increase due to absorption of starlight. Lastly, a number of groups
\citep{2002A&A...385..166S, 2003ApJ...587L.117C, 2005ApJ...618..512B,
2005ApJ...629L..45C, 2005A&A...436..719I} are beginning to model the
day-night winds on tidally locked, short orbital period
planets, and the role of clouds
\cite{2003ApJ...589..615F} . As we 
stress here, the crucial parameter for the cooling rate is the
temperature at the radiative-convective boundary, which is orders of
magnitude deeper in pressure than the photosphere.

The plan of the paper is as follows. The uncertain surface boundary
condition is discussed in \S \ \ref{sec:surfacebc},
motivating the surface isotherm used in our models. Details of cooling
models and microphysical input are described in 
\S \ \ref{sec:numerics}. In \S \ \ref{sec:luminosity} we compute the
dependence of 
the luminosity on planet mass, core entropy and irradiation.  An
analytic solution for the temperature profile in the radiative zone,
and the position of the radiative-convective boundary are
derived in \S \ \ref{sec:transition}.
These results are collected together in
\S \ \ref{sec:cooling} to derive an analytic cooling model which
exhibits simple power-law dependence on time. The radii of irradiated
gas giant planets are discussed in \S \ \ref{sec:radiusev}, and the
analytic formula for the radius given in eq.\ (\ref{eq:Rtfit}).  We
apply our models to the observed transiting planets and give
constraints on the temperature of the deep surface isotherm in \S \
\ref{sec:applications}. Our main conclusions are summarized in \S \
\ref{sec:conclusions}. 


\section{ Surface Boundary Condition }
\label{sec:surfacebc}

The surface boundary condition we adopt is to set the temperature  $T \equiv
T_{\rm deep}$ at a sufficiently large optical depth that the stellar
light is fully absorbed, and the radiation diffusion approximation is
valid. This choice of surface boundary condition has also recently
been advocated by 
\citet{2005A&A...436..719I}, based on the results of time-dependent
radiative models for the atmosphere of HD 209458b. We motivate
our choice with a simple 
toy problem, and then discuss its relation to detailed radiative
transfer solutions for the atmosphere. 

The atmosphere is heated by absorption of starlight, and possibly
dissipation of day-night winds and tidal flows. Let there be an energy
deposition rate $\varepsilon$ per unit volume in a radiative region of
thickness $\Delta z$. Choose boundary conditions $T=0$ at the top (for
simplicity) and outward flux $F=0$ at the base of the heated layer. The
latter choice is required in steady state so that the temperature deeper
in the atmosphere not increase in time. The flux generated in the layer,
which exits the planet, is $F = \varepsilon \Delta z$, and the temperature
of the deep atmosphere is $T_{\rm deep} \sim (\tau F/\sigma)^{1/4}$,
where $\tau=\kappa \rho \Delta z$ is the optical depth, $\rho$ is the
density and $\kappa$ is the opacity. Hence an atmosphere subject to
intense heating is expected to develop a deep isothermal region below the
heated layer, the temperature determined primarily by the energy flux and
depth of the layer, through $\tau$. This estimate  of the deep isotherm
temperature is similar  to that found for absorption of starlight for
the proper choice of $\tau$ \citep{2003ApJ...594.1011H}.

 We now discuss the temperature profile
for static atmospheres in more detail. In the absence of external
irradiation, the photosphere of a planet will 
cool to a temperature $T_{\rm cool} \sim  (F_{\rm  cool}/\sigma)^{1/4}
\sim 100\ {\rm K}$ in a few Gyr's, where $F_{\rm  cool}$ is the flux
from the deep interior. A characteristic temperature at small
optical depth for an
irradiated planet can be defined by balancing absorbed and emitted
energy flux.  For a star with mass $M_*$,  radius $R_*$ and effective
temperature $T_*$ a distance  $a=(GM_*P_{\rm orb}^2/4\pi^2)^{1/3}$
away, this ``equilibrium'' temperature is
\be
&& T_{\rm eq}  \equiv T_*
(R_*/2a)^{1/2} \nonumber \\ & \simeq &    1400\ {\rm K} 
\left(\frac{{\rm 3\ day}}{P_{\rm orb}}\right)^{1/3} 
\left(\frac{T_*}{6000\ {\rm K}}\right) 
\left(\frac{R_*}{R_\odot}\right)^{1/2}
\left(\frac{M_\odot}{M_*}\right)^{1/6},
\label{eq:Teq}
\ee  
an order of magnitude larger than for an isolated planet. Hence
the surface boundary condition is drastically altered from the
isolated case. In general, the irradiated boundary condition will
cause the planet to cool slower \citep{2000ApJ...534L..97B}, as we
discuss in detail. As significant horizontal temperature variation is expected
above the photosphere, $T_{\rm eq}$ is an average temperature which
gives
the correct outgoing flux.
Eq.\ (\ref{eq:Teq}) assumes zero
reflection of the stellar photons, and should be multiplied by
$(1-A)^{1/4}$ for nonzero Bond albedo $A$.

The (optical) incoming stellar photons not scattered back out of the
planet are absorbed at the starlight's photosphere, typically at a
pressure $\la 10^6\ {\rm dyne\ cm^{-2}}$.  Radiative balance implies
an outgoing (infrared) flux $F \sim (T_{\rm eq}/T_{\rm cool})^4 F_{\rm
cool} \sim 10^4 F_{\rm cool}$ generated by thermal emission. This large
flux may lead to a significant increase in temperature above $T_{\rm
eq}$ (the greenhouse effect, e.g. Hubeny et al. 2003). This situation
continues to a depth at which the starlight is fully absorbed, at which
point the temperature profile becomes isothermal. Hence, a semi-infinite
atmosphere subject to external irradiation, and with  no internal flux
deep in the atmosphere, becomes isothermal at large optical depth. We
label the temperature of this deep isotherm $\Tdeep$. Now including the
internal cooling flux $F_{\rm cool}$, the temperature will again rise
toward the interior, the gradient eventually becoming large enough for
convection to occur.

Since the cooling luminosity is generated in deep layers with
sufficiently large optical depth that the stellar light is fully
absorbed, the radiation diffusion approximation is valid
there. Furthermore, we will show in \S \ \ref{sec:transition} that the
temperature profile becomes isothermal within a pressure scale height
of the radiative-convective boundary. Hence the problem of determining
the cooling luminosity is insensitive to many of the details of the
absorption of starlight. The only input needed from the full radiation
transfer problem near the photosphere is the temperature of the deep
isotherm, $T_{\rm deep}$. \footnote{ We expect that the degree to which this
layer is isothermal depends on the number of pressure scale heights
separating the optical photosphere from the radiative-convective
boundary. Larger irradiation and lower core entropy should make this
layer more nearly isothermal.}

For tidally locked planets, the day side will be significantly hotter
than the night side in static atmospheres with negligible day-night
winds. A more uniform temperature distribution results if winds can
carry heat from the day to the night side without suffering radiative
losses (e.g. Iro et al. 2005). We will show that the
cooling luminosity is determined in deep layers with thermal time
$t_{\rm th} \ga 10^3\ {\rm yr}$. While significant day-night
temperature asymmetries may exist near the optical photosphere, winds
moving at even a tiny fraction, $\sim 10^{-5}$, of the sound speed could
deposit heat on the night side in less than a thermal time at the
depths where the cooling luminosity is determined. Hence we have a
strong expectation of a near-spherically symmetric, isothermal
temperature profile deep in the radiative layer. 


\section{ Numerical Models for the Interior }
\label{sec:numerics}

In the deep interior where the diffusion approximation is valid, we
solve the mechanical and thermal structure equations  
\citep{1959flme.book.....L}
\be
\frac{dm}{dr} & = & 4\pi r^2 \rho, \\ 
\frac{dP}{dr} & = & - \frac{Gm\rho}{r^2}, \\ 
\frac{dT}{dr} & = & \frac{dP}{dr} \frac{T}{P} \nabla, \\  
\frac{dl}{dr} & = &  \frac{dm}{dr} \left( \varepsilon - T \frac{\partial
  S}{\partial t} \right)
\label{eq:entropy},
\ee
for the interior mass $m$, pressure $P$, temperature $T$, and
outward luminosity $l$, as a function of radius $r$. Here $S$ is the entropy
per gram, 
 and $\nabla=d\ln T/d\ln P$ is the logarithmic temperature
gradient. The energy generation $\varepsilon$ is set to zero
throughout this paper, as we study passively cooling planets.
 The subscript ``cool'' on the luminosity will be assumed for
the rest of the paper.  As
the eddy turnover time is much shorter than the cooling time and
convection is quite efficient, entropy is very nearly constant in
space in the convection zone, but decreases in time due to
cooling. Hence we treat $\partial S/\partial t$ as a constant in the
convection zone. For numerical convenience, we use this same value of
$\partial S/\partial t$ in the surface radiative zone. A negligible
luminosity is generated there however, so this error does not affect
our results. While the entropy equation (\ref{eq:entropy}) is valid on
timescales longer than an eddy turnover time ($\sim {\rm yrs}$) in the
convective core, the assumption of nearly spatially constant
luminosity is only 
valid in the radiative envelope on timescales longer than the thermal
time ($\sim 10^3\ {\rm yr}$) there. As this is much shorter than the
global cooling time, we expect our numerical cooling models to be
as accurate as a relaxation (Henyey-type) code.

The equation of state (EOS) from \cite{1995ApJS...99..713S} (SCVH) is used
with a mixture of $70\%$ hydrogen and $30\%$ helium, ignoring metals, and
using the tables which smooth over the plasma  phase transition. There
are been several improvements to SCVH \citep{1999astro.ph..9168S,
2004ApJ...608.1039F} using recent laser shock- compression data and
including the effects of helium phase separation, which change the radii
at the few percent level.  We use the solar composition ``condensed''
phase opacities from \citet{2001ApJ...556..357A}, which includes the
effects of grains in the equation of state, but ignores their opacity,
as is appropriate if the grains have condensed out. Mixing length theory
is used to calculated $\nabla$  in convective regions, and radiative
diffusion in radiative regions. The mixing length is set equal to the
pressure scale height.

Two boundary conditions are needed at the surface. First, the surface
temperature is set to $\Tdeep$, the temperature of the deep isotherm
discussed in \S \ \ref{sec:surfacebc}. The second boundary
condition is that we specify the surface to be at the (arbitrarily
chosen) pressure $P=10^4\ {\rm dyne\ cm^{-2}}$. As the surface layer
is isothermal, the contribution to the radius from near-surface layers
is larger than for the radiative zero temperature profile, hence care
is needed when comparing the radii computed here with previous
work. As the radius is somewhat dependent on the problem at hand
(optical photosphere versus infrared photosphere, corrections due to
geometry in a transit, etc.) we have made this arbitrary choice of the
surface for simplicity. The change in radius  between pressures $P_1$
and $P_2$ is  $\Delta R=\int_{P_1}^{P_2} d\ln P (k_bT/\mu m_p g)
\simeq (k_bT/\mu m_p g) \ln(P_2/P_1)$. For example, the radius must
be decreased by $\Delta R=-0.022R_J$ for an outer boundary condition
$P=10^6\ {\rm dyne\ cm^{-2}}$ for $T=1000\ {\rm K}$ and mean molecular
weight $\mu=2.43$ ($70\%$ molecular hydrogen and $30\%$ neutral helium).
We do not include a solid core in the present calculations.

We make a single model of a planet as follows. Planet mass $M$, core
entropy $S$, and surface temperature $\Tdeep$ are treated as fixed
parameters. Assuming values for the planet's radius $R$, cooling
luminosity $L=l(R)$, $\partial S/\partial t$, and central pressure
$P_c$, we integrate outward from the center and inward from the
surface. The four parameters are adjusted to make the
integration variables ($m$, $P$,  $T$, and $l$) continuous at a fitting
radius.  Given the subroutine to solve for a single model, evolving
the planet in time is trivial. As we specify the core entropy $S$, and
have solved for $\partial S/\partial t$, we compute the time it
takes to cool from one entropy to the next.


\section{ Impact of irradiation on heat loss }
\label{sec:luminosity}

We now show the dependence of the cooling luminosity on the depth of
the radiative-convective boundary, emphasizing the role of the opacity
deep in the planet. Many of the luminosity dependences can be
understood with purely local arguments, without the  need to build a
global planet model. Hence, qualitative statements can be made about
cooling of EGP's under irradiation just given EOS and opacity
tables. We make comparisons between the local arguments and the global
numerical calculations as well.

The convective core is capable of transporting enormous
luminosities through fluid motion. Hence it is the large thermal
resistance of the outer radiative envelope that determines the
cooling flux. For an opacity which increases inward from the surface, this
resistance is largest at the base of the radiative layer, hence it is
the radiative-convective boundary that determines the cooling
flux. This boundary is moved to higher pressures by irradiation
\citep{1996ApJ...459L..35G}. 

The outward flux carried by radiative diffusion is
\be
F & = & - \frac{16\sigma T^3}{3\kappa \rho} \frac{dT}{dr},
\label{eq:flux}
\ee
where $\kappa$ is the Rosseland mean opacity.
The maximum flux which can be carried by radiative diffusion is found
using the adiabatic temperature gradient $dT/dr|_{\rm
  ad}=(\nabla_{\rm ad}T/P)(-Gm\rho/r^2)$,  where $\nabla_{\rm ad}=\partial
\ln T/\partial \ln P |_S$  ($=2/7$ for 
an ideal gas with five degrees of freedom) is the adiabatic
temperature gradient.
Multiplying by $4\pi r^2$,
the maximum luminosity per unit mass which can be carried by
radiative diffusion  at a local temperature $T$, pressure
$P$, opacity $\kappa(T,P)$, and enclosed mass $m \simeq M$ is
\be
\frac{L}{M} & = & \frac{64\pi G}{3} 
\frac{\sigma T^4}{\kappa P} \nabla_{\rm ad}.
\label{eq:Lmax}
\ee 
Choosing an entropy $S$, the right hand side of eq.\ (\ref{eq:Lmax}) can
be evaluated along an adiabat out from the center, yielding the
cooling flux for a specified temperature $T_{\rm rcb}$ at 
the radiative-convective boundary.  Eq.\ (\ref{eq:Lmax}) shows that the
luminosity per unit mass depends solely on the  entropy and
irradiating flux, and that the luminosity is proportional to the
planet's mass. \footnote{It is commonly stated that the luminosity {\it
decreases} with increasing mass. This is true at fixed core
temperature, rather than fixed entropy. The derivatives can be related
by $\partial \ln L/\partial \ln M|_{T_c}=\partial \ln L/\partial \ln
M|_S + \partial \ln L/\partial S|_M \partial S/\partial \ln
M|_{T_c}$. At fixed core temperature, entropy  increases for
decreasing mass (see Figure \ref{fig:Tcore_vs_M}).}
 
\begin{figure}
\plotone{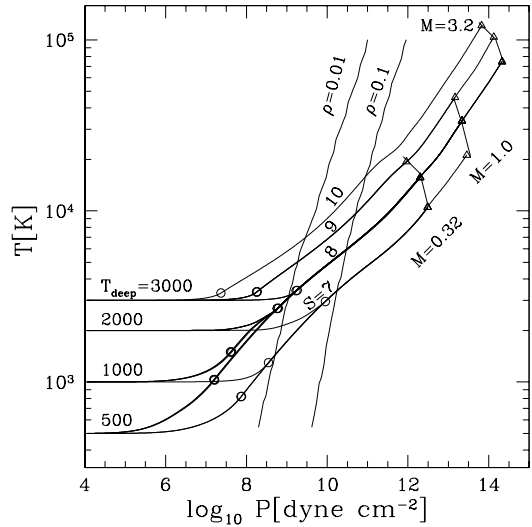}
\caption{Run of temperature vs. pressure for numerical models with
  $\Tdeep=500, 1000, 2000, 3000\ {\rm K}$, $Sm_p/k_b=7, 8, 9, 10$ and
  $M/M_J=0.32,1.0,3.2$. Only the 19 curves in the age range
  $0.1-10\ {\rm Gyr}$ are shown out of the total 48 curves. The circles
  show the position of the radiative-convective boundary. The
  triangles mark the center of the planet. Curves for
  different masses at the same $S$ and $\Tdeep$ nearly overlie each
  other. The two nearly vertical lines show contours of constant density
  $\rho=0.01\ {\rm g\ cm^{-3}}$ (left) 
  and $0.1\ {\rm g\ cm^{-3}}$ (right).
  } 
 \label{fig:T_vs_P} 
\end{figure}

Figure \ref{fig:T_vs_P} shows the run of temperature versus pressure from
numerical models for a range of $M$, $S$ and $\Tdeep$. Choosing $S$
and $\Tdeep$, the temperature profile must follow the adiabat deep in
the planet and the isotherm near the surface. An even stronger
statement can be made, however. The temperature profile over the
entire planet from the center to the top of the deep isotherm depends
only on $S$ and $\Tdeep$, and is independent of $M$ (since
eq.\ (\ref{eq:Tprofile}) and (\ref{eq:Pdeep}) depend only on $F/g
\propto L/M$). 
Next, for a given irradiation flux (fixed $\Tdeep$), the 
radiative-convective transition burrows deeper into the planet with
time (decreasing $S$). Increasing the irradiation flux at fixed $S$
also moves the radiative-convective region deeper into the
planet. 

Temperature changes in response to small changes in flux in optically thick
regions occur on the thermal time, estimated from eq.\ (\ref{eq:entropy}) to be
\be
t_{\rm th} & = & \frac{PC_pT}{gF} \simeq 10^4\ {\rm yr} 
\left(\frac{P}{10^8\ {\rm dyne\ cm^{-2}}} \right) 
\left( \frac{T}{10^3\ {\rm K}} \right)
\nonumber \\ &&
\left( \frac{10^3\ {\rm cm\ s^{-2}}}{g} \right)
\left( \frac{10^4\ {\rm erg\ cm^{-2}\ s^{-1}}}{F} \right).
\ee
Here we have used typical numbers from Figure \ref{fig:T_vs_P} for the
radiative-convective boundary. Note that this
estimate is much longer than the adjustment time near the optical
photosphere ($\sim {\rm days}$, e.g. Iro et al. 2005),
as the cooling flux is 
$\sim 10^4$ times smaller than the stellar flux, and the heat content
increases $\propto TP$. As the thermal time at the
radiative-convective boundary is so much longer than the horizontal
sound travel time ($\sim {\rm days}$), we expect the day-night
temperature asymmetry to be small there.

\begin{figure}
\plotone{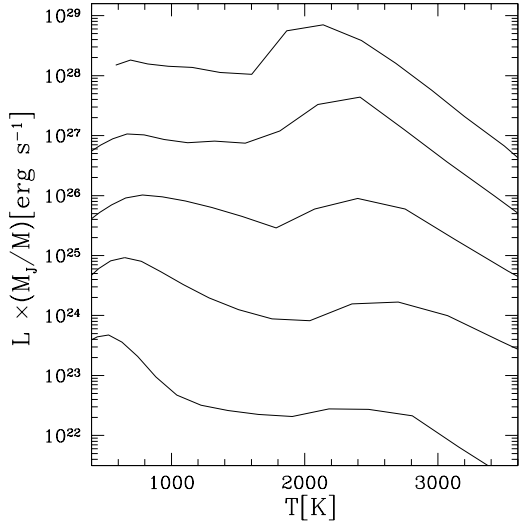}
\caption{Maximum radiative luminosity (scaled to $M=1M_J$; 
  see eq.\ [\ref{eq:Lmax}]) which can be carried by radiative diffusion
  vs. temperature along adiabats. The temperature $T$ should be
  interpreted as $T_{\rm rcb}$, the temperature at the
  radiative-convective boundary.  The five 
  curves are for entropies  $Sm_p/k_b=6,...,10$, from bottom to top.}   
 \label{fig:LoverM_vs_T}
\end{figure}

\begin{figure}
\plotone{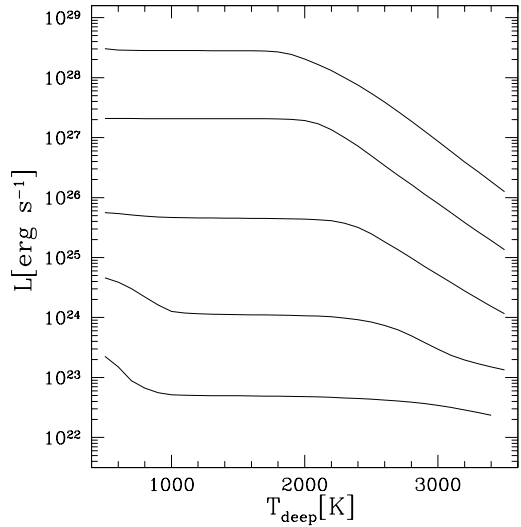}
\caption{Luminosity
  vs. temperature of the deep isotherm 
  from numerical calculation. Note that regions of positive slope in
  figure \ref{fig:LoverM_vs_T} correspond to zero slope here, implying
  the radiative-convective boundary jumps to a depth at which the slope is 
  again negative.
  Calculation is for $M=M_J$, but can be
  extended to other masses using $L \propto M$. The lines represent
  core entropies $Sm_p/k_b=6,...,10$ from bottom to top.  }  
\label{fig:LoverM_vs_T_int}
\end{figure}

Figures \ref{fig:LoverM_vs_T} and \ref{fig:LoverM_vs_T_int}
show the local calculation of $L/M$ evaluated along adiabats, and the
global calculation of $L/M$ versus $\Tdeep$, respectively. The x-axis
in Figure \ref{fig:LoverM_vs_T} is the local temperature, which should be
interpreted as $T_{\rm rcb}$, the temperature of the
radiative-convective boundary. Care must be taken in
Figure \ref{fig:LoverM_vs_T} in regions where $L(T)$ increases inward. As
we show in \S \ \ref{sec:cooling}, $L(T)$ must {\it decrease} inward in
order for convection to begin. Hence, if the chosen isotherm
intersects a region of positive slope, such as the bump near
$T=2000-2500\ {\rm K}$, the convection zone actually begins at a
deeper point at which the slope $L(T)$ is again negative. Such regions
correspond to the flat parts of the curves in
Figure \ref{fig:LoverM_vs_T_int}.  The result is that the luminosity
generally decreases with irradiation temperature, or is roughly
constant, but should not increase. This is the origin of the result
found by previous investigators \citep{2000ApJ...534L..97B} that
irradiated planets cool 
slower. Comparison of Figures \ref{fig:LoverM_vs_T} and
\ref{fig:LoverM_vs_T_int} show rough agreement in regions where $L(T)$
is decreasing, the main discrepancies due to the ratio $T_{\rm
rcb}/\Tdeep$ not being precisely a constant (see
Figure \ref{fig:T_vs_P}). 

\begin{figure}
\plotone{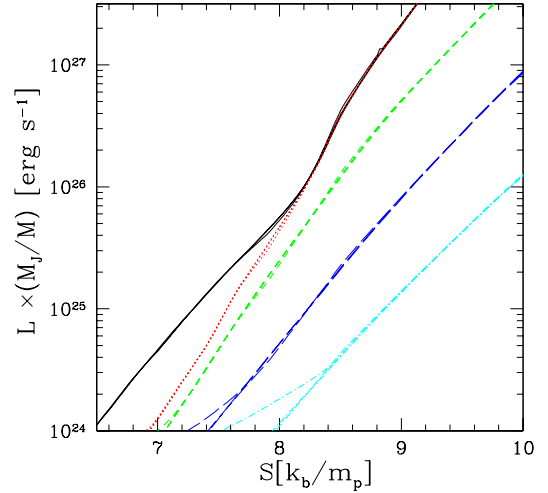}
\caption{Luminosity per unit mass (scaled to $1\ M_J$) vs. entropy
  from numerical integrations. 
  The lines correspond to surface isotherms $\Tdeep[K]=500$ (solid
  black), $1000$, (dotted red), $2500$ (short dashed green), $3000$
  (long dashed blue), $3500$ (dot short dash cyan). The lines for
  $T_{\rm deep}=1500$ and $2000\ {\rm K}$ overlie those for $T_{\rm
  deep}=1000\ {\rm K}$ (see Figure \ref{fig:LoverM_vs_T_int}). Three
  masses $M/M_J=0.32, 1.0, 3.2$ are plotted, but closely overlie each
  other (except the largest masses at 
  low entropies which are never reached) showing $L/M$ is
  independent of mass at fixed $S$ and $T_{\rm deep}$.  }   
 \label{fig:LoverM_vs_S_num}
\end{figure}

Figure \ref{fig:LoverM_vs_S_num} shows luminosity versus core entropy
for the numerical models. If $\Tdeep$ is constant during the
evolution, Figure \ref{fig:LoverM_vs_S_num} shows the change in
luminosity as the planet cools. Comparison of lines with different
$\Tdeep$ clearly shows the monotonic decrease in luminosity as the
irradiation temperature is increased. Aside from models with large mass
($M=3.2M_J$) and irradiation temperature ($\Tdeep=3500\ {\rm K}$) at
entropies so low ($S<8k_b/m_p$) as 
to be unreachable in a Hubble time, the luminosity is proportional to
the mass and the curves overlie each other. 

\section{ Radiative-convective boundary }
\label{sec:transition}

We now derive a simplified analytic model for the temperature
profile at the transition from the surface radiative zone to the core
convection zone. We relate $T_{\rm deep}$
to $T_{\rm rcb}$, the radiative-convective boundary temperature where
the cooling luminosity is determined. The
scalings derived here are used in \S \ \ref{sec:cooling} to
derive the scalings of the cooling luminosity.

\begin{figure}
\plotone{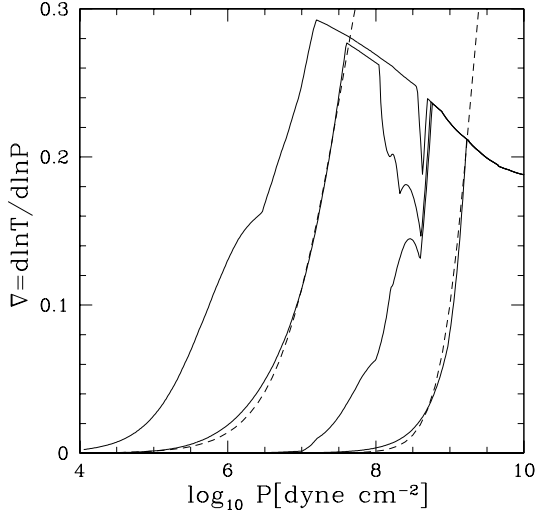}
\caption{ Solid lines show logarithmic temperature gradient $\nabla$
  as a function of 
  pressure for models with $M=M_J$, $S=8k_b/m_p$ and $T_{\rm
  deep}[K]=500, 1000, 2000, 3000$ from left to right. The two
  dashed lines 
  show the analytic formula in eq.\ (\ref{eq:nabla}) with parameters
  $(\nabla_\infty,a,P_{\rm deep}[{\rm dyne\
  cm^{-2}}])=(0.5,0.0,3.5\times 10^7)$ and $(0.5,1.0,2.0\times 10^9)$.
  The upper
  envelope is set by the adiabatic gradient $\nabla_{\rm ad}$ in the
  convection zone. Note the appearance of a radiative window which
  causes the radiative-convective boundary to be at $P \simeq
  10^{8.5}\ {\rm dyne\ cm^{-2}}$ for a large range in $\Tdeep$.}  
\label{fig:del_vs_P_allard_cond}
\end{figure}

We assume constant gravity $g$, ideal gas
pressure $P=\rho k_b T/\mu m_p$ and power-law opacity \footnote{
  Significant features in the opacity may be treated as broken power-laws.}
$\kappa\equiv\kappa_0 \rho^a T^b\equiv\kappa_1 P^a T^{b-a}$. In the radiative
zone, 
\be
F & = & \frac{16\sigma T^3 g}{3\kappa}\frac{dT}{dP}
= \frac{a+1}{4+a-b} \frac{16\sigma g}{3\kappa_1}\frac{dT^{4+a-b}}{dP^{a+1}}.
\ee
When integrating this equation, it's essential to retain the constant
of integration. Defining the temperature gradient for a radiative zero
solution $\nabla_\infty=(a+1)/(4+a-b)$, we find
\be
T^{4+a-b} & =& {\rm constant} + \nabla_\infty^{-1}
\left(\frac{3\kappa_1 F}{16\sigma g} \right)  P^{a+1}.
\ee
At small pressure, $T \simeq \Tdeep$, so we write the
temperature profile as 
\be
T & =& \Tdeep \left[ 1 + \left(P/P_{\rm deep}
  \right)^{a+1} \right]^{1/(4+a-b)},
\label{eq:Tprofile}
\ee
which becomes isothermal below the pressure
\be
P_{\rm deep} & = & \left( \nabla_\infty \frac{16 \sigma g
  \Tdeep^{4+a-b}}{3\kappa_1 F} \right)^{1/(a+1)}. 
\label{eq:Pdeep}
\ee
The logarithmic temperature gradient is then
\be
\nabla & = & \nabla_\infty \frac{ \left(P/P_{\rm deep} \right)^{a+1} }{1
  + \left(P/P_{\rm deep} \right)^{a+1} },
\label{eq:nabla}
\ee 
which decreases sharply over a pressure scale height. A plot of
$\nabla$ versus $P$ is shown in Figure \ref{fig:del_vs_P_allard_cond}
for several values of $\Tdeep$. The upper envelope of the curves is
set by the adiabatic gradient in the convection zone. Increasing
$T_{\rm deep}$ moves the boundary inward along the adiabat,
$P_{\rm deep} \propto \Tdeep^{1/\nabla_{\rm ad}}$, aside from regions
where the opacity changes irregularly. Eq.\ (\ref{eq:nabla}) agrees
well with the numerical integrations in regions where the opacity is
smooth.

To solve for the
transition from radiative to convective zone, we set
$\nabla=\nabla_{\rm ad}$. As $a+1>0$, one must have the
inequality $\nabla_\infty \geq \nabla_{\rm ad}$ for a convection zone
to exist. We find the temperature and pressure at the boundary are
\be
T_{\rm rcb} & = & \Tdeep \left(
\frac{\nabla_\infty}{\nabla_\infty-\nabla_{\rm ad}}
\right)^{1/(4+a-b)}
\label{eq:Ttr}
\nonumber \\
P_{\rm rcb} & = & P_{\rm deep} \left(
\frac{\nabla_{\rm ad}}{\nabla_\infty-\nabla_{\rm ad}} \right)^{1/(a+1)},
\ee
so they differ by a factor of order unity from $\Tdeep$ and
$P_{\rm deep}$ unless  $|\nabla_{\rm
  ad}-\nabla_\infty|<<\nabla_\infty$. The decrease of $\nabla_{\rm ad}$
at large pressures seen in Figure \ref{fig:del_vs_P_allard_cond} makes
the ratio $T_{\rm rcb}/\Tdeep$ closer to unity for large $T_{\rm deep}$
and $P$, as seen in Figure \ref{fig:T_vs_P}. Also note that at fixed $S$
and $\Tdeep$, $T_{\rm rcb}$ is largely independent of mass.


\section{ Analytic cooling model }
\label{sec:cooling}

In \S \ \ref{sec:luminosity} we found that the luminosity scales
with core entropy and irradiating flux over much of the 
relevant parameter space. Here we show that the core temperature also
scales simply with mass and entropy when sufficiently degenerate. As a
consequence, we derive an analytic model in which entropy has a
simple power-law time dependence at late times. We compare the
power-law model against numerical time integrations.

The scaling of luminosity with $\Tdeep$ and $S$
can be found by substituting $P(T,S)$ into eq.\ (\ref{eq:Lmax}). 
Using the thermodynamic relation
\be
\frac{dS}{C_p} & = & \frac{dT}{T} - \nabla_{\rm ad} \frac{dP}{P},
\label{eq:thermo}
\ee
and expanding about a reference point
$T_{\rm ref}$, $P_{\rm ref}$ and $S_{\rm ref}$, the adiabat is
\be
P & \simeq & P_{\rm ref} \left( \frac{T}{T_{\rm ref}}
\right)^{1/\nabla_{\rm ad}} \exp\left(-\frac{\Delta S}{C_p\nabla_{\rm
    ad}} \right),
\label{eq:adiabat}
\ee
where $\Delta S=S-S_{\rm ref}$, and we approximate $\nabla_{\rm
  ad}$ and $C_p$ as constants. For an ideal gas, $C_p \nabla_{\rm
  ad}=k_b/\mu m_p$, but particle interactions and molecular
dissociation reduce $C_p \nabla_{\rm ad}$ below the ideal value.
Inserting eq.\ (\ref{eq:adiabat}), (\ref{eq:Ttr}), and the power-law
form of the opacity into eq.\ (\ref{eq:Lmax}), we find 
\be
L & \simeq & L_{\rm ref} \left( \frac{T_{\rm deep}}{T_{\rm ref}}
\right)^{-\alpha} 
\exp\left[ \beta \frac{(S-S_{\rm ref})}{k_b/m_p} \right]
\label{eq:Lpl}
\ee
where the exponents are (Figures \ref{fig:LoverM_vs_T_int} and
\ref{fig:LoverM_vs_S_num}) 
\be
\alpha & \simeq & (4+a-b)\left(\frac{\nabla_\infty}{\nabla_{\rm ad}} -
1\right) \simeq 0.0-10.0,
\nonumber \\
\beta & \simeq & (a+1) \frac{k_b/m_p}{C_p \nabla_{\rm ad}} \simeq 2.5-3.5.
\ee
Examination of the exponent $\alpha$ shows that irradiation slows the
cooling, i.e. luminosity decreases as $T_{\rm deep}$ increases. The
condition $\nabla_\infty > \nabla_{\rm ad}$ is required for a core
convection zone to exist, hence $\alpha \geq 0$. Evaluation of
$\alpha$ depends on the detailed density and temperature dependence of
the opacity, which can be found in 
Figure \ref{fig:LoverM_vs_T}. Features of note are the positive slope
near $2000-3000\ {\rm K}$ at which point $\alpha$ become small, and
also the steep decrease for $T_{\rm rcb} \geq 3000\ {\rm K}$.

The exponent $\beta$ can be estimated for an ideal gas (solar mixture,
molecular hydrogen and neutral helium) and density independent opacity
to be $\beta \simeq \mu \sim 
2.4$. This ideal limit is expected for small $T_{\rm deep}$ and
hence low density. As $T_{\rm deep}$ is increased,
molecular interactions make the gas less ideal, reducing the value
of $C_p \nabla_{\rm ad}$ and increasing $\beta$. This qualitative
trend may be seen in Figure \ref{fig:LoverM_vs_S_num}.

\begin{figure}
\plotone{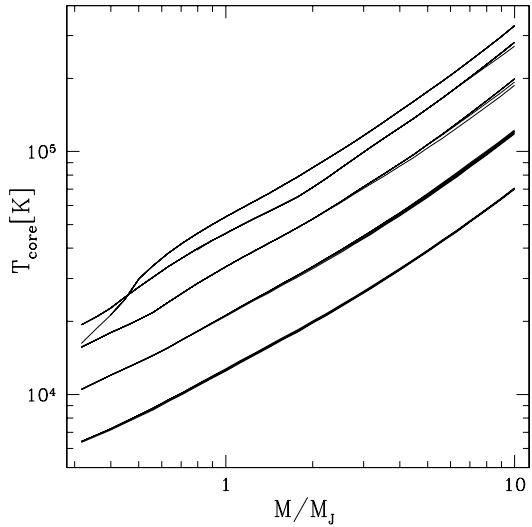}
\caption{ Core temperature vs. mass. The five groups of lines show
 entropies $Sm_p/k_b=6,...,10$ from bottom to top. Each group contains
 surface isotherms $\Tdeep[K]=500,..., 3500$, inducing larger spread
 for small entropy. }  
\label{fig:Tcore_vs_M}
\end{figure}

\begin{figure}
\plotone{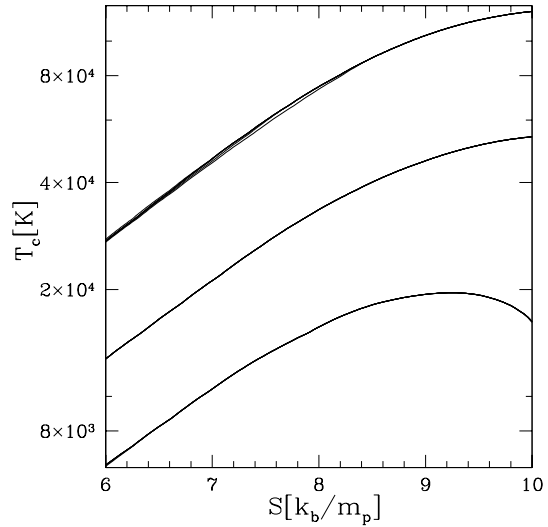}
\caption{ Core temperature vs. entropy. The three distinct lines are
  the masses $M/M_J=0.32, 1.0, 3.2$ from bottom to top. For each mass,
  the irradiation temperatures $\Tdeep[K]=500,..., 3500$ are plotted,
  but there is so little dependence on $T_{\rm deep}$ that the curves
  are indistinguishable. }  
\label{fig:Tc_vs_S}
\end{figure}

The core temperature increases during the initial contraction phase
when the core is non-degenerate. A maximum is reached when $k_bT_c
\simeq E_F$, the Fermi energy, and subsequently $T_c$ decreases as
entropy decreases.  In this degenerate phase, the core temperature
depends mainly on mass and entropy, with only a weak dependence on
irradiation. Figure \ref{fig:Tcore_vs_M} shows core temperature versus mass
for four adiabats and a range of irradiation temperatures. The
dependence on $\Tdeep$ gives only a slight broadening of each
adiabat. The dependence on mass is quite simple when sufficiently
degenerate. Figure \ref{fig:Tc_vs_S} shows the dependence of core
temperature on entropy for a range of masses and irradiation
temperatures, showing a simple exponential dependence at low
entropy. For the degenerate phase we write $T_c$ in the form (see
Figure \ref{fig:Tcore_vs_M})
\be
T_c(M,S) & = & T_{c,\rm ref}\left( \frac{M}{M_{\rm ref}}\right)^\gamma
\exp\left[ \delta \frac{(S-S_{\rm ref})}{k_b/m_p}\right].
\label{eq:Tcpl}
\ee
Using hydrostatic balance $P
\propto M^2/R^4$, and parameterizing $R \propto M^\lambda$, we estimate
the exponents to be 
\be
\gamma & \simeq & \nabla_{\rm ad}(2-4\lambda)\simeq 0.6-0.7
\nonumber \\
\delta & \simeq & k_b/C_pm_p\simeq 0.5.
\ee

Next we solve for the change in core entropy with time for the
analytic model. We treat $T_{\rm deep}$ and $M$ as constants
during the evolution. The entropy equation integrated over the convective
core gives 
\be
\frac{\partial S}{\partial t} & = &  - \frac{L/M}{fT_c},
\label{eq:intentropy}
\ee
where $f=\int (dm/M)(T/T_c)\simeq 0.6-0.7$ and we have treated $\partial
S/\partial t$ as constant in space. Plugging eq.\ (\ref{eq:Tcpl}) and
(\ref{eq:Lpl}) into eq.\ (\ref{eq:intentropy}), we find the following
solution for the entropy with time
\be
\exp\left[\frac{S-S_{\rm ref}}{k_b/m_p} \right] & = & \left( 1 +
\frac{t}{t_{\rm S}} \right)^{-1/(\beta-\delta)},
\label{eq:Svst}
\ee
where the characteristic cooling time is
\be
&& t_{\rm S}(M,T_{\rm deep}) \nonumber \\
 & = & \left( \frac{f}{\beta - \delta } \right)
\left( \frac{ k_b T_c/m_p}{ L/M } \right)_{\rm ref} 
\left( \frac{T_{\rm deep}}{T_{\rm deep,ref}}
 \right)^\alpha \left( \frac{M}{M_{\rm ref}} \right)^\gamma.
\label{eq:tS}
\ee

This solution has a number of notable features:
\begin{list}{}
\item (1) The fiducial evolution time, $Mk_b T_c/m_p L$, is just the
  {\it initial} time to radiate away the core's thermal energy. 
\item (2) At late times, $t \geq t_S$, the solution is a
  power-law in time. As the initial cooling
  time is very short, the power-law occurs for most of the
  planet's lifetime.
\item (3) The exponent of the power-law involves the change of luminosity
  and core temperature with respect to entropy.
\item (4) Evolution timescale is slowed for large irradiation or large planet
  mass. The dependence on planet mass comes purely from the dependence
  of $T_c$ on planet mass (at fixed entropy). The dependence on
  irradiation is primarily through the luminosity.
\end{list}

\begin{figure}
\plotone{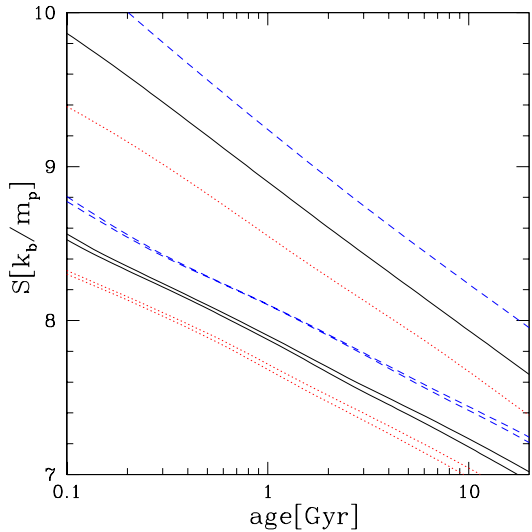}
\caption{Core entropy vs. age for a planet mass $M/M_J=0.32$ (dotted red),
 $1.0$ (solid black), and $3.2$ (dashed blue). For each mass, surface isotherms
 $\Tdeep[K]=1000,2000,3000$ are shown from bottom to top. Luminosity
 is independent of $T_{\rm deep}$ for $\Tdeep=1000-2000\ {\rm K}$,
 hence the curves nearly overlie each other (see
 Figure \ref{fig:LoverM_vs_T_int}).}  
\label{fig:S_vs_time}
\end{figure}

Figure \ref{fig:S_vs_time} shows entropy versus time for a range of mass
and irradiation. Note the large spread in $S$ at a given age. For low
irradiation, $S \simeq 6-8k_b/m_p$ in  the age range $1-10\ {\rm
Gyr}$, while $S$ can be as high as $\simeq 10k_b/m_p$ for
$\Tdeep=3500\ {\rm K}$. While the range of $\Tdeep$ shown here gives
fairly good power-laws, we note that at $\Tdeep<500\ {\rm K}$ there is
a break occurring at $\simeq 1\ {\rm Gyr}$, due to the increase in
luminosity seen in Figure \ref{fig:LoverM_vs_S_num} below $S=8k_b/m_p$.

While we have used the SCVH EOS and Allard et.al.(2001) opacities for
numerical estimates, the analytic solution makes it particularly clear
which quantities need by evaluated for a given opacity table and
EOS. The luminosity is sensitive only to the local conditions at the
radiative-convective boundary, while the core temperature involves
building static (i.e. not time-dependent) models.


\section{ Radius Evolution }
\label{sec:radiusev}

Planets are initially nondegenerate in their core, and undergo rapid
contraction until the core is 
degenerate. If $T_{\rm eff}$  is
constant during 
the contraction, the energy equation $d/dt(-3GM^2/7R)=-4\pi R^2 \sigma
T_{\rm eff}^4$ is solved to find the change in radius with
time (see, e.g. Bildsten et al. 1997)
\be
R(t) & \simeq & 8\ R_J\ \left( \frac{M}{M_J} \right)^{2/3} \left(
\frac{300\ {\rm K}}{T_{\rm eff}} \right)^{4/3} \left( \frac{1\ {\rm
    Myr}}{t} \right)^{1/3}.
\ee
The core temperature $T_c \simeq GM\mu m_p/Rk_b \propto t^{1/3}$ is
increasing during the non-degenerate phase, and reaches a maximum when
$k_bT \simeq E_F$. For an ideal gas,
$k_bT/E_F$ is a function only of entropy, so that the maximum
temperature would occur at the same entropy for all planet
masses. Coulomb interactions suppress the
value of $\nabla_{\rm ad}$ below $2/5$, so that if $T_c \propto
M^{4/3}$ and $P_c \propto M^{10/3}$, the entropy at maximum
temperature will increase a bit with mass. This can be seen in
Figure \ref{fig:Tc_vs_S}, as the two higher masses have maxima at higher
entropy (off the plot) than the lowest mass.

\begin{figure}
\plotone{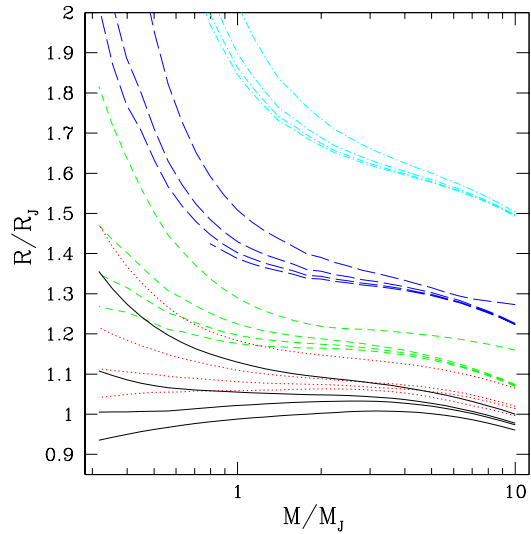}
\caption{Radius vs. mass curves for different core entropy and irradiation.
  Each group of lines
  with a different color and line style denote different entropies.
  Solid black, dotted
  red, and short-dashed green, long dashed blue and dot-dashed cyan 
  lines represent entropies $Sm_p/k_b=6,...,10$. Each group of lines
  represents deep isotherms $T_{\rm deep}=500, 1500, 2500,3500\ {\rm K}$,
  from bottom to top.  } 
 \label{fig:R_vs_M_irr}
\end{figure}

Once degeneracy sets in, the radius is primarily determined in the
degenerate core of the planet, although as irradiation is increased
the contribution from the outer
envelope becomes larger due to the increased scale height. This is
clarified by writing the radius as an integral over pressure
\be
r(P) & = &  \int^{P_c}_P d\ln P \left(  \frac{P}{\rho g} \right) .
\ee
In the degenerate core, $P/\rho \propto  E_F$ while in the
nondegenerate envelope $P/\rho  \propto k_bT$. The
contribution from the core is larger when $E_F \gg k_b T$ unless the
number of pressure scale heights in the envelope is much larger than
the core. 

The equation of state in the core for $M<M_J$ is
complicated by strong Coulomb interactions. For illustrative purposes,
an approximate equation of
state including the leading order contributions from Coulomb
interactions as well as ideal ion pressure is
\be
P & = & n_e \left( \frac{2}{5} E_F - \frac{3}{10} \frac{Z^2e^2}{a_i} \right) +
\frac{\rho k_b T}{\mu_i m_p},
\label{eq:simpleeos}
\ee
where $a_i=(4\pi \rho/\mu_i m_p)^{1/3}$ is the mean ion spacing and 
$\mu_i m_p$ is the mean ion mass. The energy scales relevant for the
core are $E_F=(\hbar^2/2m_e)(3\pi^2
\rho/\mu_e m_p)^{2/3} \simeq 26\ {\rm eV}(\rho\mu_e^{-1}{\rm g\
  cm^{-3}})^{2/3}$, $k_b T 
\simeq 0.9\ {\rm eV} (T/10^4\ {\rm K})$, and $E_{\rm coul}=Z^2e^2/a_i
\simeq 20\ {\rm eV} Z^2  (\rho\mu_i{\rm g\ cm^{-3}})^{1/3}$ is the Coulomb
interaction energy between a nucleus and uniform electron cloud.
Ignoring the ion pressure term, the density 
at zero pressure is $\rho_{\rm zp}\simeq 0.2 \mu_e Z^2
{\rm g\ cm^{-3}}$. As the central density in Jupiter mass objects is
near $\rho_{\rm zp}$, Coulomb interactions (and further the tendency
to form bound states) stiffen the EOS, and are important in
determining the radius.

\begin{figure}
\plotone{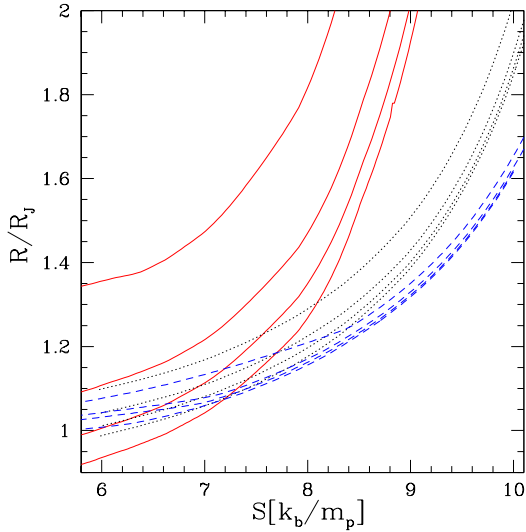}
\caption{Radius vs. core entropy for different masses and irradiation. 
  Solid red, dotted black, and dashed
  blue lines represent masses $M/M_J=0.32,1.0,3.2$. Each group of lines
  represents $T_{\rm deep}=500, 1500, 2500, 3500\ {\rm K}$, from bottom to
  top. } 
 \label{fig:R_vs_S}
\end{figure}

Figure \ref{fig:R_vs_M_irr} shows mass versus radius for a range of core
entropy and irradiation. The effects of irradiation are seen to be
most severe at low mass and low entropy, since $\Tdeep$ is becoming a
significant fraction of the core temperature. At $M \simeq M_J/2$ and
low entropy, the range of irradiation temperatures shown here can 
change the radius by as much as $50\%$. Radii for fully adiabatic
planets (not shown here) agree well with the $\Tdeep=500\ {\rm K}$
lines.

Figure \ref{fig:R_vs_S} shows radius versus entropy for a range of
masses and irradiation temperature. At late times in the evolution
when the entropy  is small, the radius is converging to some constant
value which depends on {\it both} $M$ and $\Tdeep$. If the planet were
allowed to cool under a constant irradiation field indefinitely, it
would approach an isothermal state \citep{1977Icar...30..305H} at
$T=\Tdeep$ with a radius \footnote{In principle, $R_0$ can be
calculated by integrating the structure equations for a given  EOS. In
practice, such low temperatures and high densities are not covered by
the SCVH EOS. In this paper, we compute the isothermal radius by
fitting evolutionary curves of radius versus entropy, defining $R_0$
by extrapolating to the small entropy limit.}  $R=R_0$. Although in
practice planets will  never reach this isothermal state, it is the
{\it minimum} radius to which the planet is evolving. Furthermore, it
is the {\it deviation} around the isothermal radius, $\delta R=R-R_0$
which is changing with age. As we now show, $\delta R$ has a
particularly simple behavior with time over the entire observable
range $\delta R \leq R$.

To motivate the following numerical calculations, we first discuss the
change in radius for a fluid element in mass shell $m$ as the
entropy is changed. The radius of a mass shell in the convection
zone can be written
\be
r^3(m,S) & = & \frac{3}{4\pi} \int_0^m \frac{dm'}{\rho(m',S)},
\ee
hence for fixed interior mass the change in radius with respect to
entropy is
\be
\frac{\partial r}{\partial S} & = & - \frac{1}{4\pi r^2}
\int_0^m \frac{dm'}{\rho(m',S)} \frac{\partial \rho(m',S)}{\partial
  S} \rfloor_{m'}. 
\ee
Given an equation of state $\rho(P,S)$, and switching to radius as the
integration variable, we find
\be
\frac{\partial r}{\partial S} & = & - \frac{1}{r^2}
\int_0^r {r'}^2 dr' \left( \frac{1}{C_p}
\frac{\partial \ln \rho}{\partial \ln T}\rfloor_{P}
+ \Gamma_1^{-1} \frac{\partial \ln P}{\partial S}\rfloor_{m'} \right)
\label{eq:drds}
\ee
where eq.\ (\ref{eq:thermo}) has been used. The second term in
eq.\ (\ref{eq:drds}) mainly corresponds to a uniform shift in pressure in
the core, due to the radius changing. Near the surface this term
must go to zero since pressure is proportional to external mass, which
is fixed. Consequently, the first term is most
important. From eq.\ (\ref{eq:simpleeos}), the volume expansion term is 
$\partial \ln \rho /\partial \ln T|_P \propto k_bT/E_F$, with a
significant correction due to Coulomb interactions which acts to
increase the expansion since the electron pressure is effectively
lowered. Hence the change in radius in the core is proportional
\footnote{ Eq.\ (\ref{eq:simpleeos}) has ignored contributions to the
  Coulomb correction which depend on temperature, and do not scale
  linearly with temperature. Using the EOS in
  \citet{2000PhRvE..62.8554P}, we find the contribution of these terms
  to the 
  volume expansion seems to be somewhat smaller than the ideal ion
  pressure. } to
$T_c$. As $T_c$ depends exponentially on the entropy
(eq.\ [\ref{eq:Tcpl}]), the contribution to the radius from the degenerate
core depends exponentially on entropy. In the nondegenerate envelope,
$\partial \ln \rho /\partial \ln T|_P \simeq -1$. Plugging this result
into eq.\ (\ref{eq:drds}) implies that the change in radius due to the
nondegenerate envelope scales linearly with entropy. As a consequence,
it is less important than the exponential dependence from the core.

\begin{figure}
\plotone{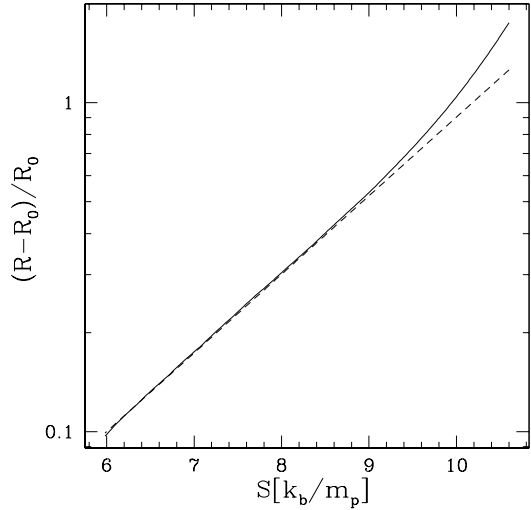}
\caption{Comparison of computed fractional radius $(R(S)-R_0)/R_0$
  (solid line) to fitting formula $R(S)=R_0+\delta R_S e^{\eta m_p
  S/k_b}$ (dashed 
  line) for a $M=1\ M_J$ planet with $T_{\rm deep}=1000\ {\rm K}$.}   
 \label{fig:goodfitexample}
\end{figure}

A suite of evolutionary calculations has been done for
$M/M_J=0.32,1.0,3.2$ and $\Tdeep[K]=500, 1000, ..., 3500$ starting
from high entropy and evolved to ages greater than $15\ {\rm
  Gyr}$. Given the run of $R(S)$, we fit a function
\be
R(S) & = & R_0+\delta R_0\exp(\eta m_p S/k_b)
\label{eq:Rfit}
\ee
to determine the isothermal radius $R_0$, coefficient $\delta
R_0$,
and exponent $\eta$. The coefficients $R_0$, $\delta R_0$ and $\eta$
depend 
on $M$ and $\Tdeep$. The small 
entropy points were more heavily weighted to force the fit to agree
there. The weighting was adjusted until the fit agreed  for
as large a region in $S$ as possible (for the plots here we used
weighting $\propto R^{10}$.) A comparison of the fit against the data
for one example is given in Figure \ref{fig:goodfitexample}. The
agreement is good at small entropies, and gets worse for large entropy
as degeneracy is lifted. We find good agreement between $\eta \sim
0.5-0.7$ and $\delta$ from eq.\ (\ref{eq:Tcpl}), as expected if
$\delta R \propto  T_c$.

\begin{figure}
\plotone{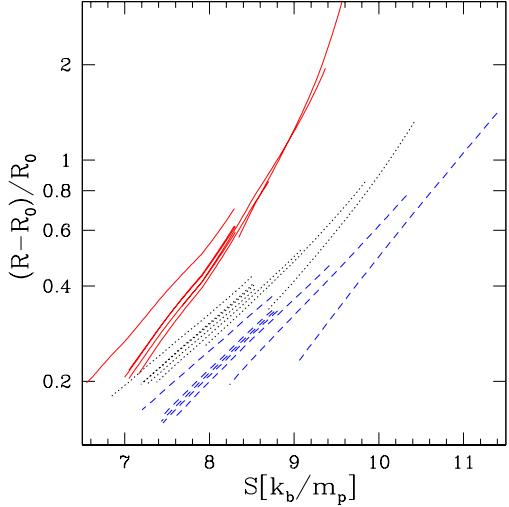}
\caption{Fractional deviation in radius vs. core entropy for different
masses and irradiation. This curves give a sense of how much the radius
changes during evolution. Solid red,
  dotted black, and dashed blue 
  lines represent masses $M/M_J=0.5,1.0,1.5$. Each group of lines
  represents $T_{\rm deep}=500,1000,...,3500\ {\rm K}$, from left to
  right. Only ages in the range $0.1$ to $10\ {\rm Gyr}$ are shown for
  each curve. }  
 \label{fig:dR_vs_S}
\end{figure}

The deviation of the radius about the isothermal value is plotted for
all runs over the age range $0.1-10\ {\rm Gyr}$ in
Figure \ref{fig:dR_vs_S}. Recall that $R_0$ is different for each
line. Note that each line is approximately a power-law, even to
$\delta R/R_0 \simeq 1$, where the degenerate approximation breaks
down. Hence the fitting formula often works better than naively
expected.

\begin{deluxetable}{rrrrr}
\tablecolumns{5}
\tablewidth{0pc}
\tablecaption{Parameters for the Fitting Function $R(t)$ in
  eq.\ (\ref{eq:Rtfit}). }
\tablehead{
\colhead{$M/M_J$} & \colhead{$\Tdeep [K]$} &
\colhead{$\eta/(\beta-\delta)$} & \colhead{$R_0/R_J$} &
\colhead{$\delta R_1/R_J$} } 
\startdata
   0.316   & 500   & 0.31 &   0.836  & 0.308 \\
   0.316   &  1000 &  0.25 &  0.881 & 0.330 \\
   0.316  & 1500   & 0.23   & 0.905  & 0.359 \\
   0.316  & 2000   & 0.23   & 0.952 & 0.361 \\
   0.316  & 2500   & 0.32   & 1.05 & 0.401 \\
   0.316  & 3000   & 0.43   & 1.15 & 0.867 \\
   0.316  & 3500   & 0.50  & 1.30 & 2.49 \\
   1.00  & 500 & 0.16   & 0.825 & 0.320 \\
   1.00  & 1000   & 0.16  & 0.894 & 0.273 \\
   1.00  & 1500   & 0.15   & 0.902 & 0.280 \\
   1.00  & 2000  & 0.16   & 0.930 & 0.266 \\
   1.00  & 2500 & 0.24  & 0.992 & 0.264 \\
   1.00  & 3000   & 0.25   & 0.993 & 0.439 \\
   1.00 & 3500  & 0.35  & 1.09 & 0.727 \\
   3.16  & 500  & 0.20 & 0.915 & 0.243 \\
   3.16  & 1000   & 0.16  & 0.934 & 0.238 \\
   3.16  & 1500   & 0.17   & 0.947 & 0.227 \\
   3.16  & 2000   & 0.18   & 0.962 & 0.217 \\
   3.16  & 2500   & 0.26 & 0.994 & 0.239 \\
   3.16 & 3000 & 0.30 & 1.02 & 0.379 \\
   3.16  & 3500  & 0.41   & 1.10 & 0.673 \\
\enddata
\label{tab:fits}
\end{deluxetable}

\begin{figure}
\plotone{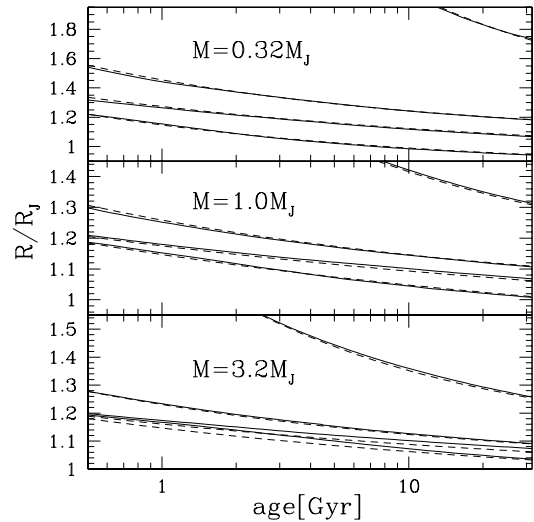}
\caption{Radius vs. age for $M/M_J=0.32, 1.0, 3.2$ and $\Tdeep=500, 1500,
  2500,3500\ {\rm K}$  from bottom to
  top within each panel. Solid lines show the numerical evolutions,
  while dashed lines show the fitting formula in eq.\ (\ref{eq:Rtfit}).}
 \label{fig:R_vs_age}
\end{figure}

We now combine the power-law cooling result in eq.\ (\ref{eq:Svst}) and
(\ref{eq:tS}) with the fit for the radius in eq.\ (\ref{eq:Rfit})
to find
\be
R(t) & = & R_0 + \delta R_0 \exp(\eta m_pS_{\rm ref}/k_b) \left( 1 +
\frac{t}{t_{\rm S}} \right)^{-\eta/(\beta-\delta)}.
\ee
At late times $t \gg t_S$, the deviation in radius from the isothermal
value is a power-law in time. In order to provide useful fits to our
evolutionary tracks, we parametrize this late time power-law as
\be
R(t) & = & 
R_0 + \delta R_1 \left(
\frac{{\rm 1\ Gyr}}{t} \right)^{\eta/(\beta-\delta)},
\label{eq:Rtfit}
\ee 
where $R_0$ is again the isothermal radius and $\delta R_1=\delta R_0
\exp(\eta m_pS_{\rm ref}/k_b) (t_{\rm S}/{\rm 1\
  Gyr})^{\eta/(\beta-\delta)}$ is the 
deviation at an age of $1\ {\rm Gyr}$. We fit tracks of $R(t)$ to find
the coefficients $R_0$, $\delta R_1$ and $\eta/(\beta-\delta)$ in the
same way as the fits for $R(S)$ in eq.\ (\ref{eq:Rfit}). The coefficients
are given in Table \ref{tab:fits}. Comparison between the numerical
evolutionary tracks for $R(t)$ and the analytic fit in
eq.\ (\ref{eq:Rtfit}) are given in Figure \ref{fig:R_vs_age}.
The agreement is generally very good.

Approximate values and scalings of the coefficients in Table
\ref{tab:fits} can be understood as follows.  The expected power-law
index $\eta/(\beta-\delta) \simeq 0.6/3.0 = 0.20$ agrees well with the
temperature range $\Tdeep=1000-2000\  {\rm K}$ where $L$ is
independent of $\Tdeep$. At large irradiation, Figure \ref{fig:dR_vs_S}
shows $\eta$ increases and Figure \ref{fig:LoverM_vs_S_num} shows that
$\beta$ decreases, explaining the increase in
$\eta/(\beta-\delta)$. Since $\delta R_1 \propto
t_S^{\eta/(\beta-\delta)}
\propto \Tdeep^{\alpha \eta/(\beta-\delta)}$, regions of constant
(decreasing) slope in Figure \ref{fig:LoverM_vs_T_int} correspond to
$\delta R_1$ being constant (increasing). The magnitude of $\delta
R_1$ can be estimated from Figure \ref{fig:dR_vs_S} and
eq.\ (\ref{eq:tS}). Interestingly, $R_0$ can be somewhat bigger for
$M=0.32M_J$ than for the higher masses. While a larger radius is
expected for strong irradiation, we caution the reader about
interpretation of the exact values for $R_0$. It would be interesting
to compare the values obtained by fitting tracks with actual
calculations of isothermal planets given a sufficiently accurate low
temperature EOS.

\begin{figure}
\plotone{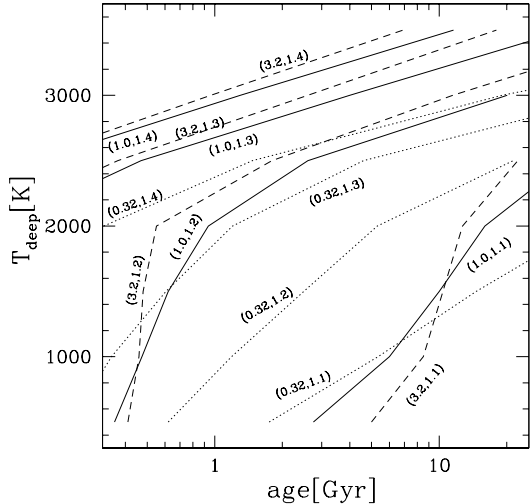}
\caption{$\Tdeep$ vs. age for planets with a given $M$ and $R$. Shown
  are lines of constant radius for a given mass, labeled by
  $(M/M_J,R/R_J)$. At a given age, this plot shows the value of
  $\Tdeep$ required to explain a certain mass and radius. There is
  significant degeneracy between $\Tdeep$ and age.  }
 \label{fig:Tdeep_vs_age}
\end{figure}

Given measurements of planetary mass, radius and age, $T_{\rm deep}$
can be constrained. Figure \ref{fig:Tdeep_vs_age} shows the value of
$T_{\rm deep}$ required to explain a planet of a given mass and radius,
as represented by different lines, as a function of age. The lines slope
up to the right since the cooling must be slower (higher $T_{\rm deep}$)
to reach the same radius at larger age. As the lines are not horizontal
or vertical, there is significant degeneracy between $T_{\rm deep}$
and age. Large radii in the age range $1-10\ {\rm Gyr}$ can only be
explained by large irradiation temperatures for the mass range $0.32-3.2
M_J$. For each mass and radius, there is a minimum age which is set
by the unirradiated planet, resulting in a steep slope down to the left.
We shall use Figure \ref{fig:Tdeep_vs_age} in \S \
\ref{sec:applications} to constrain $T_{\rm deep}$ for the observed
transiting planets.



\section{Applications to Transiting Planets}
\label{sec:applications}


\begin{figure}
\plotone{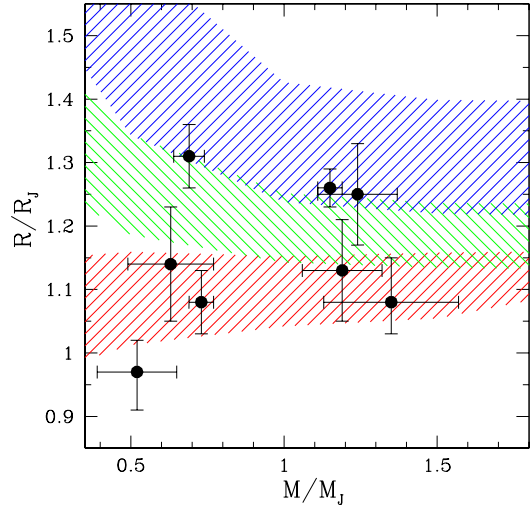}
\caption{Radius vs. mass for observed transiting planets compared to
  our cooling model. The 
  points with error bars give the observed masses and radii, as listed
  in Table \ref{tab1}. The lower (red), middle (green), and upper
  (blue) hatched areas denote $\Tdeep=500, 2500, 3000\ {\rm K}$. For
  each $\Tdeep$ and $M$,  the spread in $R$ denotes ages $1\ {\rm
  Gyr}$ (top) to $10\ {\rm Gyr}$ (bottom) in each hatched region. HD
  149026's small radius places it well outside the plot. }
 \label{fig:R_vs_M_data}
\end{figure}

We now compare our theory to the observed masses and radii of
the transiting planets (Table \ref{tab1}). Figure
\ref{fig:R_vs_M_data} shows radius versus mass for 
the observed transiting planets. The points with errorbars are the
data. The three different hatched regions show $T_{\rm deep}=500, 2500,
3000\ {\rm K}$ from bottom to top. The change is gradual from $T_{\rm
deep}=500$ to $2500$, and then accelerates for higher temperatures (see
Figure \ref{fig:LoverM_vs_T_int}). Within each hatched region, a spread of
ages from $1$ (top) to $10\ {\rm Gyr}$ (bottom) is shown. The radius of
HD 149026 is so small as to be well outside the plot. It clearly has a
large abundance of heavy elements. The radii of the other eight planets
can be broadly explained with solar composition, ages in the range $1-10\
{\rm Gyr}$, and temperatures deep in the atmosphere  $T_{\rm deep}
\leq 3000\ {\rm K}$. The largest radii requiring the most irradiation
are HD 209458, HD 189733 and OGLE-TR-56.

There are significant uncertainties in fitting
stars on the main sequence to find stellar ages. Hence there is
motivation to understand how a range of  ages affects the range of
observed radii. Figure \ref{fig:dR_vs_S} shows deviation from the
isothermal radius by factors $1.1-2$ in the age range $0.1-10\ {\rm
Gyr}$. The length of each track gives an idea of the uncertainty in
radius due to an uncertainty in age. Using the fitting formula in eq.\
(\ref{eq:Rtfit}), the fractional difference in radius between ages
$t_1$ and $t_2$  is $\simeq [\eta/(\beta-\delta)](\delta
R_1/R_0)\ln(t_2/t_1)$. For the strongly irradiated case, if we choose
characteristic values $\eta/(\beta-\delta)=0.5$, $\delta R_1/R_0=0.5$,
and a factor of two error in age $t_2=2t_1$, the fractional error in
radius is $17\%$. Hence, the age dependence for strongly irradiated
planets is important because (i) the decrease in time is steep, and
(ii) strong irradiation increases the size of $\delta R_1$ relative to
$\delta R_0$.

\begin{figure}
\plotone{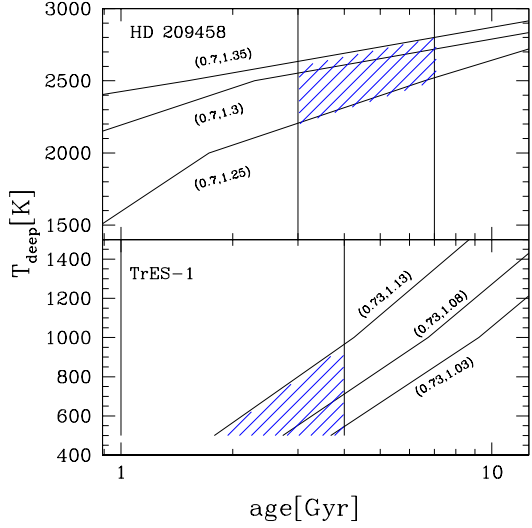}
\caption{ Constraints on $\Tdeep$ for HD 209458 and TrES-1. Lines
  labeled by $(M/M_J,R/R_J)$
  given allowed values of $\Tdeep$ versus age for the central value of
  mass, and a range of radii given by the radius error bar. Vertical
  lines give age range from main sequence fitting of parent
  star. Shaded region shows range of $\Tdeep$ allowed given
  uncertainties in mass, radius and age.  }
\label{fig:HD209458TRES1}
\end{figure}

\begin{figure}
\plotone{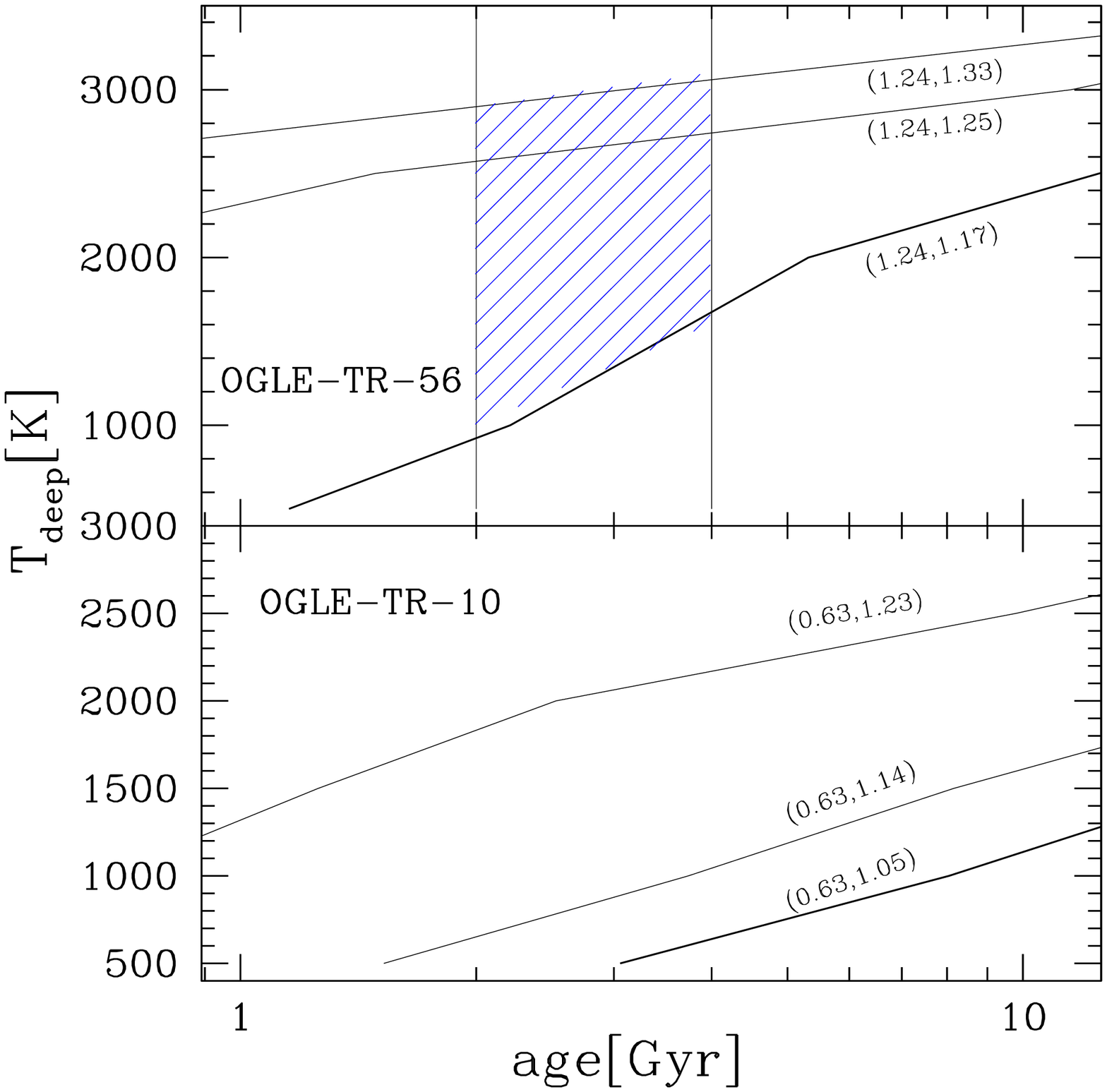}
\caption{ Constraints on $\Tdeep$ for OGLE-TR-56 and OGLE-TR-10. See
  Figure \ref{fig:HD209458TRES1} for description. }
\label{fig:OGLETR56OGLETR10}
\end{figure}

\begin{figure}
\plotone{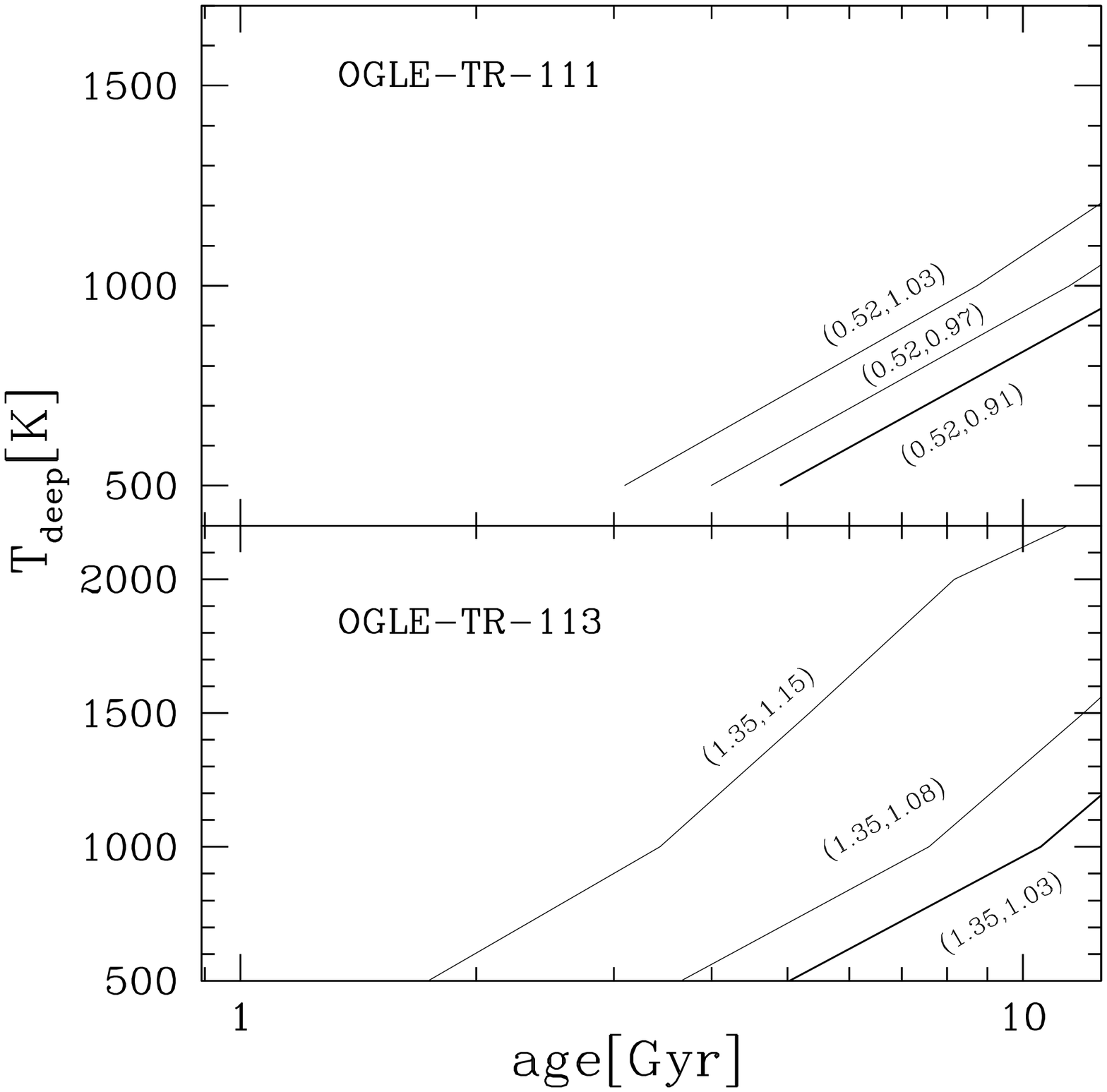}
\caption{ Constraints on $\Tdeep$ for OGLE-TR-111 and OGLE-TR-113. See
  Figure \ref{fig:HD209458TRES1} for description. }
\label{fig:OGLETR111OGLETR113}
\end{figure}

\begin{figure}
\plotone{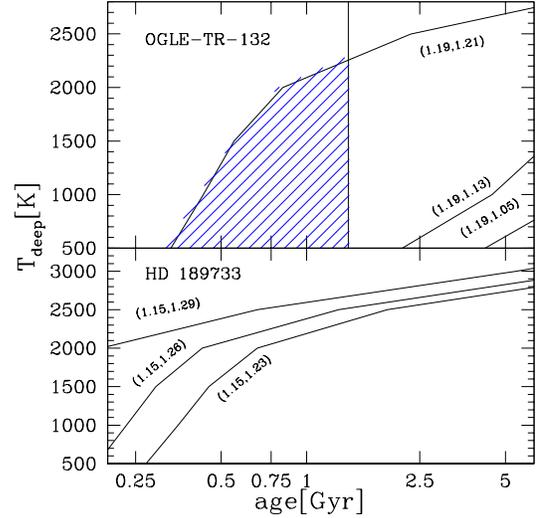}
\caption{ Constraints on $\Tdeep$ for OGLE-TR-132 and HD 189733. See
  Figure \ref{fig:HD209458TRES1} for description. }
\label{fig:OGLETR132HD189733}
\end{figure}

Next, the parameter $T_{\rm deep}$ is crucial for the cooling rate,
but is not directly measurable. Here we constrain $T_{\rm deep}$
using measured mass, radius and age. We then compare $T_{\rm
deep}$ to the equilibrium temperature.

We interpolate over the ${\rm age}-T_{\rm deep}$ tracks in Figure
\ref{fig:Tdeep_vs_age} for the mass and radii appropriate for each
planet (except HD 149026, which we do not discuss). Since the
uncertainty in $T_{\rm deep}$ due to the error bar in planet mass is
smaller than that due to the error bar in radius, we fix the
planet mass at the central value and only vary the radius. For those
planets with an age range quoted in the literature, we show the age
range in the plot, and derive the range of $T_{\rm deep}$ consistent
with the age range. These values are listed in Table \ref{tab1}.  For
those planets with no age determination, we find the maximum value of
$T_{\rm deep}$ consistent with an age less than $10\ {\rm Gyr}$.

The constraints on $T_{\rm deep}$ for all planets except HD 149026 are
shown in Figures \ref{fig:HD209458TRES1}, \ref{fig:OGLETR56OGLETR10},
\ref{fig:OGLETR111OGLETR113} and \ref{fig:OGLETR132HD189733}, and
summarized in Table \ref{tab1}.  The $T_{\rm deep}$ of HD 209458 is
best constrained due to the small error bar on mass and radius, as
well as detailed fitting of the parent star to find the age. From
Figure \ref{fig:HD209458TRES1} we find  $T_{\rm deep}=2200-2800\ {\rm
K}$; HD 209458b is not consistent with an un-irradiated
planet. OGLE-TR-56 has a weak lower limit on $T_{\rm deep}$ which is far
less than the equilibrium temperature. All
other planets with age constraints have only upper limits, set by the
upper limit on the radius, since the lower limit on the radius is
consistent with no irradiation. Plots are provided for those planets
with no age constraints at the present time.

\begin{figure}
\plotone{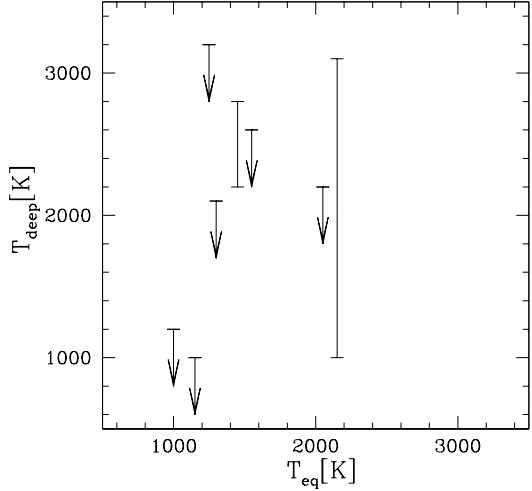}
\caption{ Equilibrium temperature vs. $T_{\rm deep}$. The temperature of
the deep isotherm $T_{\rm deep}$ is expected to be strongly correlated
with the equilibrium temperature $T_{\rm eq}$, giving a line sloping up
to the right. Bracketed lines give constraints on $T_{\rm deep}$, 
while arrows pointing
downward represent upper limits found when no age was available in the
literature.}
\label{fig:TeqvsTdeep}
\end{figure}

The equilibrium temperature  $T_{\rm eq} \equiv
T_{\ast}(R_\ast/2a)^{1/2}$  is measurable, but plays no part in our
model. On the other hand, the temperature of the deep isotherm is not
measurable, but is crucial for the cooling rate. From radiative
transfer models, we expect these two temperatures to be roughly
proportional, the exact ratio determined by the size of the greenhouse
effect (\S \ref{sec:surfacebc}). Hence, they should be strongly
correlated.  Figure \ref{fig:TeqvsTdeep} shows a plot of $T_{\rm eq}$
versus $T_{\rm deep}$. The large error bars on $T_{\rm deep}$, due to
large error bars on the radius, prevent one from drawing robust
conclusions. It is in principle possible for a correlation (sloping up 
to the right) to exist given the current error bars, however, it is
not required.  

Tighter constraints on $T_{\rm deep}$ require the following:
\begin{itemize}
\item Significantly smaller error bars on the radii of OGLE-TR-56 and
OGLE-TR-132.
\item An age estimate is needed for HD 189733. If it is found to have an age
  $\geq 1\ {\rm Gyr}$, $T_{\rm deep}$ will be well constrained with a
  value much larger than $T_{\rm eq}$, similar to HD 209458b.
\item Age estimates are needed for OGLE-TR-10, OGLE-TR-111, and
  OGLE-TR-113. However, given the present error bar on radii of
  OGLE-TR-10 and OGLE-TR-113, $T_{\rm deep}$ will be constrained only
  at the factor of two level. OGLE-TR-111 is an interesting case, as
  it must be older than $\sim 3-4\ {\rm Gyr}$ to be consistent with our model.
\end{itemize}
We encourage efforts in these directions.

\section{conclusions}

\label{sec:conclusions}

We have presented calculations of cooling and radius
evolution for strongly irradiated planets. Novel aspects of this model
are the following:

\begin{itemize}

\item We argue that the generic outcome of strong surface heating, whether it
  be due to absorption of stellar flux or dissipation of winds and
  tidal flow, is that a deep isothermal region exists above the
  radiative-convective boundary. The
  thermal time in this layer is sufficiently long that the temperature
  profile is approximately spherically symmetric, irrespective of the
  size of the asymmetry near the photosphere. We assign this region
  the temperature $T_{\rm deep}$ and treat it as a boundary
  condition for the cooling models.

\item We show that the cooling flux is determined at the
  radiative-convective boundary, which is much deeper than the
  photosphere. Scalings of the flux with
  core entropy, $T_{\rm deep}$, and mass are computed.

\item These scalings allow us to derive an analytic model for the cooling,
  which shows power-law decrease over a large range of parameter
  space. The part of the radius which changes in time (the deviation from
  the isothermal planet) is also a
  power-law. An analytic formula for radius evolution is given in
  eq.\ (\ref{eq:Rtfit}), with coefficients in Table \ref{tab:fits}.

\item While we have used the SCVH EOS and Allard et.al.(2001)
  opacities for numerical estimates, the analytic solution makes it
  particularly clear  which quantities need to be evaluated for a given
  opacity table and EOS. The luminosity is sensitive only to the local
  conditions at the radiative-convective boundary, while the core
  temperature involves building static (i.e. not time-dependent)
  models.


\end{itemize}

We have compared our theory to observed masses and radii for the
transiting planets in Table \ref{tab1} (except for HD 149026, which
clearly has a large abundance of heavy elements). Our findings are as follows:
\begin{itemize}

\item Figure \ref{fig:R_vs_M_data} shows mass versus radius for eight
  transiting planets, compared to our model. The radii can be broadly
  explained with solar composition, ages in the range $1-10\ {\rm
  Gyr}$, and temperatures deep in the atmosphere  $T_{\rm deep} \leq
  3000\ {\rm K}$. The largest radii requiring the most irradiation to
  explain are HD 209458, HD 189733 and OGLE-TR-56.

\item Figures \ref{fig:HD209458TRES1}, \ref{fig:OGLETR56OGLETR10},
  \ref{fig:OGLETR111OGLETR113} and \ref{fig:OGLETR132HD189733} show
  constraints on $T_{\rm deep}$ using measured masses, radii, and ages
  (when available), and their uncertainties. We find that only HD
  209458b is well constrained, with $T_{\rm deep}=2200-2800\ {\rm
  K}$. OGLE-TR-56 has a weak lower limit, and the other six planets
  have only upper limits, due to the
  large measurement uncertainty in the radius, or lack of an age
  determination. These constraints are summarized in Table \ref{tab1}.

\item The equilibrium temperature $T_{\rm eq}$ is measurable, but
  plays no part in our model. The deep isothermal temperature $T_{\rm
  deep}$ is not measurable, but is crucial for the cooling
  rate. Radiative transfer calculations find these two temperatures
  should be strongly correlated. Figure \ref{fig:TeqvsTdeep} shows
  $T_{\rm eq}$ versus $T_{\rm deep}$. As only upper limits on $T_{\rm
  deep}$ are available for all but HD 209458b and OGLE-TR-56, 
  it is difficult to draw
  conclusions at the present time.  It is in principle possible  for a
  correlation to exist given the current error bars, however, it is
  not required.

\end{itemize}

We hope that our models have illuminated the need for more accurate
ages and radii. Once those are in hand, our calculations
will provide a measurement of $T_{\rm deep}$ of adequate accuracy to
compare to $T_{\rm eq}$, thus constraining greenhouse physics and
day-night transport.


\vspace{4cm}

\acknowledgements 

This project arose out of a lunchtime conversation with Adam Burrows
discussing cooling models for gas giant planets.  We thank Tristan
Guillot and France Allard for helpful advice on opacities. We also
thank Omer Blaes, Shane Davis, Eric Pfahl and Evan Scannapieco for
useful discussions. We would also like to thank the referee for constructive
comments which improved the presentation of this paper. Phil Arras was
supported by the NSF Astronomy and Astrophysics Postdoctoral Fellowship,
and the Kavli Institute for Theoretical Physics during this project.
This work was supported by the National  Science Foundation  under
grants PHY99-07949 and AST02-05956.



\end{document}